\documentclass[twocolumn,showpacs,amsmath,amssymb,prb]{revtex4}  

\usepackage{graphicx}

\begin{document}
\title{Generalized method of image and the tunneling spectroscopy in high-$T_{c}$
superconductors }
\author{Shin-Tza Wu$^{1}$ and Chung-Yu Mou$^{1,2}$}
\affiliation{
1. Department of Physics, National Tsing Hua University, Hsinchu 30043,
Taiwan\\
2. National Center for Theoretical Sciences, P.O.Box 2-131, Hsinchu, Taiwan}
\date{\today}

\begin{abstract}
A generalized method of image is developed to investigate 
the tunneling spectrum from the metal into a class of states,
with the tight-binding dispersion fully included.
The broken reflection symmetry is shown to be
the necessary condition for the
appearance of the zero-bias
conductance peak (ZBCP).
Applying this method to the $d$-wave superconductor yields results
in agreement with experiments regarding
the splitting of ZBCPs in magnetic field. Furthermore, a ZBCP is
predicted for
tunneling into the (110) direction of the $d$-density wave state, 
providing a signature to look for in experiments.
\end{abstract}
\maketitle

The current transport through a heterojunction consisting of a normal metal
and another different material ($X$) has been the subject of interest for many
years. In this setup, the normal metal with known spectral properties is
used as a probe to analyze the electronic states of the material $X$.
\cite{wolf} Although such measurements have provided useful
insights into the bulk spectral properties of $X$, it has been also
realized that the presence of the interface matters.  The zero-bias
conductance peak (ZBCP) observed in the tunneling spectra when $X$ is a
$d$-wave superconductor (ND junction) in (110) direction\cite{Hu1} is a
well-known example of interface effects. However, the issue of exactly
how the tunneling measurements are related to the bulk properties has
never been answered satisfactorily.  Conventionally, the ND junction
is analyzed in the mean-field level, using the Bogoliubov de Gennes
(BdG) equations in which continuum and quasi-classical approximations
are often invoked.  While these approximations are valid for
conventional superconductors, they are certainly not justified for
high Tc cuprates where proximity to the Mott insulators entails fully
consideration of the tight-binding nature.  Previously\cite{Tanuma}
this was done by numerically solving the discrete BdG equation for
each interface orientation individually without elucidating their
relations to the bulk properties. This technical inconvenience makes
it difficult to include fluctuations systematically in this approach.

In this work, we shall adopt a different approach based on the
non-equilibrium Keldysh-Green's function formalism which enables one
to construct systematically higher order corrections from the mean
field lattice Green's functions.\cite{mou1,Yeyati,Luck}  In this approach,
because $X$ extends over a semi-infinite space, one shall need the
half-space Green's functions.  For simple configurations such as the
(100) orientation of a $d$-wave superconductor, it turns out that this
half-space Green's functions only differ from the bulk ones by
sinusoidal factors. This relation certainly does not hold for other
orientations as it predicts no ZBCP in the (110) direction.
We shall develop a generalized method of image which enables us to
construct the half-space Green's function from the bulk ones. {\em We
emphasize the generality of this method and its ability to 
account for the low energy features in the tunneling spectrum for a
whole class of states}. As a demonstration, in this article we will
focus mostly on the study of ND junctions and only briefly mention the
applications to other systems. The effects of interactions and 
fluctuations will be addressed elsewhere. 

Our results indicate broken reflection symmetry is necessary for
the emergence of ZBCP. For ND junctions our method can reproduce 
earlier results on the ZBCP in the continuous wave 
approximation.\cite{Hu1,datta} In a full tight-binding
calculation for (110) and (210) directions, we obtain the doping dependence of
the ZBCPs which shows its sensitivity to the Fermi surface
topology.  In particular, the splitting of the ZBCP in the 
current-carrying state is also calculated and is shown to be in agreement with
experiments. At the end, we analyze the case when $X$ is the $d$-density
wave (DDW) state in (110) direction and the semi-infinite graphene
sheet with a zig-zag type interface. The former state was recently
proposed as a possible normal state for high Tc
cuprates.\cite{Laughlin} Conductance peaks are found for both states.

We start by considering a junction consisting of a 2D normal metal on the
left ($L$) hand side ($-\infty <x\leq -d$, $d$ is the lattice constant of
the metal side) and a $d$-wave superconductor 
($0\leq x<\infty $) on the right ($R$%
) hand side (see Fig.~1), governed by the Hamiltonian $H_{L}$ and $H_{R}$
respectively. At the mean-field level, we have
\begin{eqnarray}
H_{R}&=&-\!\!\! \sum_{<ij>,\sigma} \!\!\! t_{R}c_{i\sigma }^{+}c_{j\sigma}
-\!\!\! \sum_{<ij>^{\prime },\sigma }\!\!\! t_{R}^{\prime }\,c_{i\sigma }^{+}c_{j\sigma}
\nonumber\\
& & +\sum_{<ij>}\Delta _{ij}
(c_{i\uparrow }c_{j\downarrow }-c_{i\downarrow}c_{j\uparrow })
+{\rm H.c.} \,,  \label{Hm}
\end{eqnarray}
where $<ij>$ denotes the nearest-neighbor (NN) bond, $<ij>^{\prime }$ the
next NN bond, and $\Delta _{ij}$ possesses the $d$-wave symmetry. 
The tunneling Hamiltonian connects the interface points at $%
x=-d$ and $x=0$, and is given by $H_{T}=\sum_{y}t(|y_{L}-y_{R}|)(c_{L}^{%
\dagger }c_{R}+c_{R}^{\dagger }c_{L})$, where the summation is over lattice
points along the interface, chosen to be in the $y$ direction. We shall assume
that both sides are square lattices and characterize the orientation of RHS
by the Miller indices ($hk0$). The total grand Hamiltonian is then given by $%
K=(H_{L}-\mu _{L}N_{L})+(H_{R}-\mu _{R}N_{R})+H_{T}\equiv K_{0}+H_{T}.$ Here $%
\mu _{L}$ and $\mu _{R}$ are the chemical potentials and their difference $%
\mu _{L}-\mu _{R}$ is fixed to be the voltage drop $eV$ across the junction .
\begin{figure}
\includegraphics*[width=80mm]{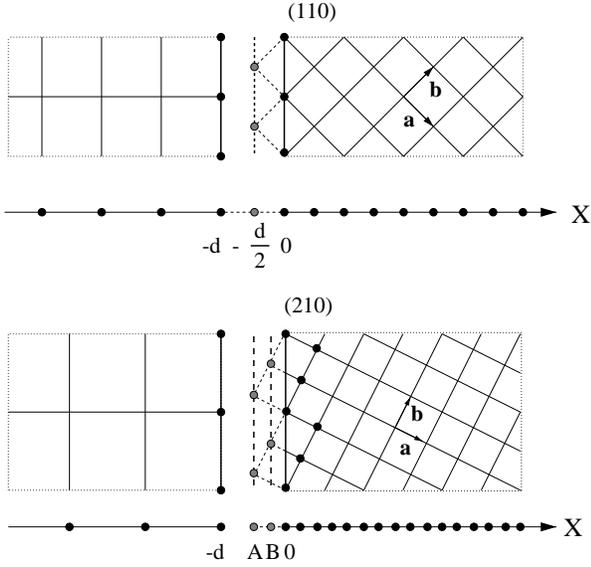}
\caption{\small Schematic plot of the ND junction in (110) and (210) directions. Here
two hard walls for the (210) case are at $A=-2d/\sqrt{5}$ and $B=-2d/\sqrt{5}
$. The dots on the $x$ axis are the reduced 1D lattice points.}
\end{figure}

In the non-equilibrium Keldysh-Green's function formalism, $H_{T}$ is
adiabatically turned on.\cite{mou1,Yeyati} As a result, the bare
Green's function is defined only on a half plane. Since the 
nearest-neighbor bonds to the interface sites 
are cut, there is effectively a hard wall located at say, for the (110) 
interface, $x=-d/2$. This hard-wall boundary condition prompts the 
application of the method of image. However, because lattice points 
in the half-plane usually do
not form a simple Bravais lattice and the $d$-wave gap changes sign
under reflection, the implementation of the conventional method of
image appears problematic. To overcome these difficulties, a Fourier
transform in the $y$ direction is performed first. Consider the case of 
(110) orientation, the Hamiltonian with only NN hopping becomes
\begin{eqnarray}
H_{R} &=&\!\!\sum_{i,\sigma,k_y}-2t_{R}\cos\!\left(\frac{k_{y}d}{2}\right)
c_{i\sigma }^{\dagger }(k_{y})c_{i+1\sigma }(k_{y})  
\nonumber\\
&+&\!\!\sum_{i,k_y}2i\Delta _{0}\sin\!\left(\frac{k_{y}d}{2}\right)
\left[ c_{i\uparrow }(k_{y})c_{i+1\downarrow }(-k_{y}) 
\right. \nonumber\\ && \left. \hspace*{20mm}
+c_{i\downarrow}(-k_{y})c_{i+1\uparrow }(k_{y})\right] 
+{\rm H.c.}\, ,  \label{oneD}
\end{eqnarray}
where $-\pi /d<k_{y}\leq \pi /d$ and $2\Delta _{0}$ is the gap value. The
whole problem is now one dimensional, and the hard wall becomes a point.
Note that the suppression of the gap near the interface can be taken into
account by adding self-consistent $\delta \Delta _{0}(i)$ to Eq.~(\ref{oneD}%
), and can be treated perturbatively later. In the presence of $t_R'$, 
additional terms $\sum_{i,\sigma,k_y}-2t_{R}^{\prime }\cos
\left( k_{y}d\right) \;c_{i\sigma }^{\dagger }(k_{y})c_{i\sigma
}(k_{y})+t_{R}^{\prime }\;c_{i\sigma }^{\dagger }(k_{y})c_{i+2\sigma
}(k_{y})+{\rm H.c.}$ appear. Since at the boundary, both $t_{R}$ and $%
t_{R}^{\prime }$ are cut and become dangling bonds, one needs to introduce
two hard walls at $x=-d/2$ \ and $x=-d$. For clear presentation, we shall
first set $t_{R}^{\prime }=0$. In this case, we are looking for the Green's
function, $\bar{G}_{0}(\omega ,k_{y,}x,x^{\prime })$ (which is a $2\times 2$
matrix in Nambu's notations), that satisfies the boundary condition $\bar{G}%
_{0}=0$ at $x=-d/2$. We shall suppress the dependence on $\omega $ and $%
k_{y} $. Here since $x^{\prime }$ is the location of the point source and
its image point is at $\ -d-x^{\prime }$,\cite{image} the method of image can be
employed by constructing 
\begin{equation}
\bar{G}_{0}(x,x^{\prime })=G_{0}(x-x^{\prime })-G_{0}(x+d+x^{\prime
})\,\alpha (x^{\prime }),  \label{Greal}
\end{equation}
where $G_{0}$ is the bulk bare Green's function and $\alpha $ is a matrix to
be determined. The first term is the direct propagation from $x^{\prime }$
to $x$, while the second term will reduce to the propagation from the image
point to $x$ in special situations (see below). In fact, since $\bar{G}_{0}$
has to vanish at $x=-d/2$, we obtain $\alpha (x^{\prime
})=G_{0}^{-1}(d/2+x^{\prime })G_{0}(-d/2-x^{\prime })$. Therefore, the
second term describes the propagation from $x^{\prime }$ to $x$ via the
reflection of the hard wall. The matrix $\alpha$, 
apart from fitting the boundary
condition, carries the important information about the gap
structure along the reflected path from $x^{\prime }$ to $x$. 
Note that in calculating the tunneling current, 
since $H_{T}$ only connects points
along the interface, only the surface Green's function
$\bar{g}_{0}(\omega ,k_{y})\equiv \bar{G}_{0}(x=0,x^{\prime
}=0)$ is needed.\cite{mou1,datta1} 
Writing $G_{0}$
in the Fourier $k_{x}$ space, we find
\begin{equation}
\bar{g}_{0}(\omega ,k_{y})\equiv \int_{-2\pi /d}^{2\pi /d}\frac{dk_{x}}{2\pi 
}\,G_{0}(\omega ,k_{y,}k_{x})\left[ 1-\exp (ik_{x}d)\,\alpha _{0}\right], 
\label{surfaceG}
\end{equation}
where the factor $\alpha _{0} = G_{0}^{-1}(d/2)G_{0}(-d/2)$ 
is independent of $k_{x}$. 
If the reflection symmetry holds for the state $X$, 
such as $d$-wave superconductors in (100) direction 
(in this case, $d/2$ is replaced by $d$), one has $
\alpha _{0}=1$ and Eq.~(\ref{surfaceG}) reduces
to the familiar form\cite{mou1} $\bar{g}_{0}(\omega ,k_{y})\equiv
\int_{-2\pi /d}^{2\pi /d}dk_{x}/\pi \,G_{0}(\omega ,k_{y,}k_{x})\sin
^{2}\left( k_{x}d/2\right) $. Therefore, apart from modifications due 
to the sinusoidal factors, the density of state almost has the same feature as
the bulk one. However, for other orientations such as
the (110) direction, reflection symmetry with respect to the 
interface is broken. As a result, $\alpha_0$ is not the identity matrix and 
as we shall see, this will give rise to the ZBCP.

The advantage of Eq.~(\ref{surfaceG}) is that it is purely based on the bulk
Green's functions. The interface orientation is encoded in $k_{x}$ and $%
k_{y} $. In other words, $G_{0}(\omega ,k_{y},k_{x})$, which appears in Eq.~(%
\ref{surfaceG}) and $\alpha _{0}$, is simply the usual bulk BCS
Green's function but with ${\bf k}$ being rotated by 45$^\circ$. This
technical merit is retained for other interface orientations with
${\bf k}$ being rotated by the angle in accordance with the interface
orientation.  More importantly, this also offers a scheme 
for studying fluctuations and interactions. Essentially one can take
into account these effects through the bulk Green's function
$G_0$. This will be explained in more detail in a separate
publication.

When evaluating $G_{0}(x-x^{\prime })$, the dominant
contributions come from the poles determined by $(\omega +i\eta
)^{2}-E_{k}^{2}=0$, where $E_{k}=\sqrt{%
\epsilon _{k}^{2}+\Delta _{k}^{2}}$. In the continuum limit, 
the dispersion becomes $\epsilon _{k}=\hbar ^{2} (k_{x}^{2}-k_{Fx}^{2}) /2m$ 
and $\Delta
_{k}=\Delta _{0}\cos 2(\theta -\theta _{0})$, where $%
k_{Fx}^{2}=k_{F}^{2}-k_{y}^{2}$, $\theta =\sin ^{-1}(k_{y}/k_{F})$, and $%
\theta _{0}$ is the angle between the crystal $a$ axis and $x$ direction.
At the same time, the integration range of $k_{x}$ is extended to $\pm
\infty $. There are four poles located at $\pm k_{\pm }$ with $k_{\pm
}\equiv \sqrt{k_{Fx}^{2}\pm 2m\sqrt{(\omega +i\eta )^{2}-|\Delta _{\pm }|^{2}%
}/\hbar ^{2}}$, representing particles and holes along different directions.
Here $\Delta _{\pm }$ are gaps in directions $\pm \theta $. By
contour integration, one obtains $G_{0}(x-x^{\prime })$ and thus 
$\alpha(x^{\prime })$. After some algebra and assuming that $k_{F}$ 
is large,\cite{mou2} indeed Eq.(\ref{Greal}) reproduces results obtained in 
Ref.~\onlinecite{datta} by direct solving the equations of motion.
In our approach, the continuum approximation is not necessary. To
investigate any effect that is due to the tight binding nature, the
full tight binding dispersion has to be retained. In this case, the
integration over $k_x$ can not be extended to $\pm \infty$, and thus 
poles are in different structure and a substantial
difference from the continuum approximation could result.

We now include the hopping $t_{R}^{\prime }$ for the (110) direction. 
The main complication is to
add a second hard wall at $x=-d$. This is a simple generalization of the
single hard-wall problem. One simply requires $\bar{G}_{0}$ vanish on all
these hard walls simultaneously. Therefore, we write 
\begin{eqnarray}
\bar{G}_{0}(x,x^{\prime })=
G_{0}(x-x^{\prime }) &-&G_{0}(x-x_{1}^{\prime })\alpha_{1}(x^{\prime })
\nonumber \\
&-&G_{0}(x-x_{2}^{\prime })\alpha _{2}(x^{\prime }),
\label{twowalls}
\end{eqnarray}
where $x_{1}^{\prime }=-x^{\prime }-d$ and $x_{2}^{\prime }=-x^{\prime }-2d$
are images of $x^{\prime }$. The boundary conditions at $x=-d$ and $x=-2d$
determines $\alpha _{1}$ and $\alpha _{2}$. 
The surface Green's function thus obtained is the bare one and
will get renormalized by $H_{T}$, giving rise to four different components
in the differential conductance\cite{mou1,mou2}. The strength of $H_{T}$,
characterized by $t$, determines the relative weight among each component.
In Fig.~2, we show our results for the spectrum of the total differential
conductance for various dopings. The parameters adopted are determined 
self-consistently from the mean-field slave-boson theory for the 
$t$-$t'$-$J$ model.\cite{mou1} It is clear that the 
ZBCPs are the most important features
at low energies.\cite{Hu1} Since the
ZBCP arises from the existence of zero-energy states, it must appear as
poles at zero energy in the Green's function. For (110) direction without $%
t_{R}^{\prime }$, this is entirely determined by zeros of the denominator
in $\alpha$: 
\begin{equation}
\beta (\omega ,k_{y})=\det [G_{0}(d/2)].  \label{det}
\end{equation}
In the continuum limit ($d\rightarrow 0$), $\beta $ can be evaluated
analytically: $\beta \approx -[\omega ^{2}/\bar{\Delta}%
^{2}+(\Delta _{+}/|\Delta _{+}|+\Delta _{-}/|\Delta _{-}|)^{2}]$ with $\bar{%
\Delta}=|\Delta _{+}||\Delta _{-}|/(|\Delta _{+}|+|\Delta _{-}|)$.
Therefore, poles of $\bar{g}_{0}$ at $\omega =0$ depends crucially on
whether there is a sign change of the gap on the Fermi surface. This
criterion, however, does not hold as one goes to the tight binding limit
because pairing no longer only occurs on the Fermi surface. As a result, a
numerical computation of $\beta $ is necessary. Our results indicate that
ZBCPs are sensitive to the Fermi surface topology. In fact, for (110)
surface, the height of the peak depends on the volume of the Fermi surface.
It reaches maximum when $\mu _{R}=0$ and decreases when $\mu _{R}\neq 0$.
For other orientations, the ZBCP could even disappear.

For general orientations, there could be more than one hard wall. As a
demonstration, we consider the $(210)$ interface. In this case, when $%
t_{R}^{\prime }=0$, two hard walls are located at $x=-d/\sqrt{5}$ and $-2d/%
\sqrt{5}$, in analogy to the (110) surface with $t_{R}^{\prime }$. 
A typical result for small scale of $t$ is shown in 
the inset of Fig.~2. As $
t $ increases, the zero-energy states are able to leak out, and thus the
ZBCPs get broadened. Note that lattice points with dangling bonds form a
pair-breaking region near the interface, resulting in the peaks around $
2\Delta _{0}$. They are due to quasi-particle bound states with non-zero
energy.\cite{Barash}
\begin{figure}
\includegraphics*[width=75mm]{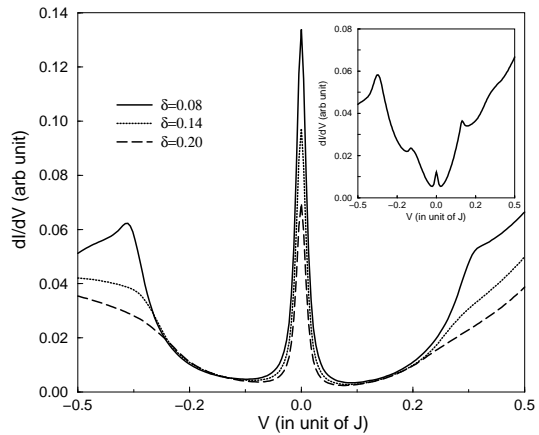}
\caption{\small The total differential conductance of several dopings for (110)
interface with $\eta=0.01$ and $t_L =1.0$. 
The weak link is modeled by the interface hopping $t(\omega )=$ $%
\exp (-\sqrt{(\omega _{0}-|\omega |)/\Gamma })$ with $\omega
_{0}=11\Delta _{0}$ and $\Gamma =\Delta _{0}$. Inset: $dI/dV$ curve for
(210) interface with $\delta = 0.08$.}
\end{figure}

The ZBCPs split in the presence of magnetic fields $H$, essentially due to
the Doppler shift caused by the supercurrent near the 
interface.\cite{Fogelstrom} In the tight-binding model, $\Delta _{ij}$ is shifted to $%
\Delta _{ij}\exp [i{\bf q}\cdot {\bf (r}_{i}+{\bf r}_{j})]$, where $%
q=eH\lambda /2\hbar c$ with $\lambda $ being the penetration depth. \ By
redefining $c_{i\sigma }=c_{i\sigma }\exp (i{\bf q\cdot r}_{i})$, the
dependence on ${\bf q}$ can be absorbed into $\epsilon _{k}$.
Figure 3 shows the field dependence of the splitting for the (110)
interface, in agreement to recent experimental data\cite{Deustch}. It 
is seen that for large $H$, the splitting deviates from linear dependence on 
$H $ due to the lattice effect in our approach.
The inset shows the doping dependence of splitting, reflecting
its sensitive dependence on the Fermi surface topology. In fact, in the
special case when particle-hole symmetry holds (for example, $\mu _{R}=0$
and $t_{R}^{\prime }=0$), we find that no matter how large $H$ is, the ZBCP
does not split at all, in consistent with naive expectations.\cite{mou2}
\begin{figure}
\includegraphics*[width=75mm]{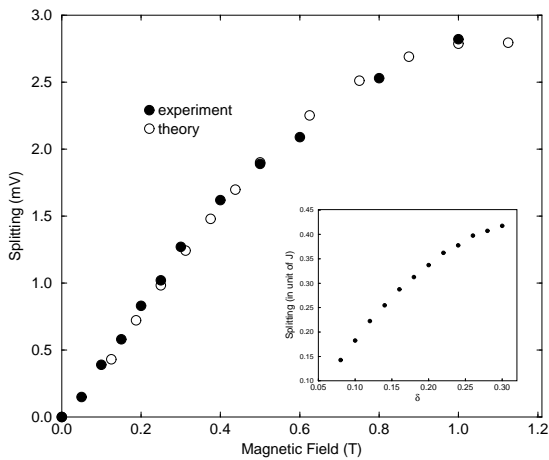}
\caption{\small  The field dependence of the splitting for an underdoped case ($\delta
=0.12)$. Inset: Doping dependence of the splitting for a fixed magnetic
field.}
\end{figure}
\begin{figure}
\includegraphics*[width=75mm]{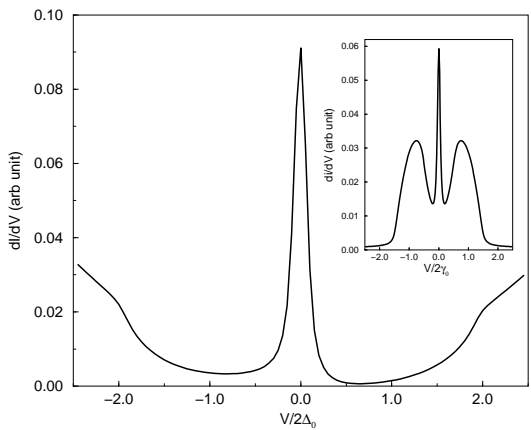}
\caption{\small A typical $dI/dV$ curve
for tunneling into the (110) direction of a $d$-density wave
state. Here $\mu _{R}=0$ and $\Delta _{0}=0.1$. Inset: A similar plot for
tunneling from a wide-band metal into the graphene sheet with zig-zag 
interface. Here the hopping amplitude $\gamma_0 = 0.1$}
\end{figure}

Finally, our approach is easily modified to deal with other states.
When $X$ is the DDW, the two-component indices are associated with the
two sublattices in this state.  The formulation presented for the ND
case can then be applied with minor modification.\cite{mou2} At (110)
orientation, the DDW state does not possess reflection symmetry; one thus 
expects ZBCP's in this case. Figure 4 shows a typical result for
tunneling into the DDW state in the (110) direction when $\mu_{R}=0$. 
Invariably, the ZBCP is present, consistent with a recent
report.\cite{sigrist} For finite $\mu_{R}$, unlike the $d$-wave
superconductor, one simply adds $\mu_{R}$ to the quasi-particle
energy $E_{k}$,\cite{Laughlin} resulting in shifting of the ZBCP to the
bias at $\mu_{R}$. The existence of this conductance peak thus
provides a signature of the DDW state.  A similar analysis can be done
for the semi-infinite graphene sheet.  When projected onto 1D lattices, 
it is then obvious that reflection symmetry is preserved in the case of 
the armchair interface while not in the zigzag case. This results in
for the latter a ZBCP in the $dI/dV$ curve (see the inset of
Fig.~4)(Ref.~\onlinecite{mou2}) -- consistent with previous numerical
work.\cite{Fujita} In conclusion, we have developed a versatile formulation 
for studying the tunneling spectra of junction systems. In particular, it 
is capable of predicting whether the ZBCP shall arise or not in the 
conductance curve. The results discussed in this article 
represent some typical examples; further applications to other systems will 
be reported elsewhere.

It is our pleasure to thank Professor Sungkit Yip, Professor Hsiu-Hau
Lin and Professor T. K. Lee for useful discussions. This research was
supported by NSC of Taiwan.


\begin{thebibliography}{99}
\bibitem{image} In fact, for 1D, the image point can be chosen arbitrarily 
without affecting the result.
\end{thebibliography}
\end{document}